# Unveil Compressed Sensing

Xiteng Liu
QualVisual Technology
Toronto, Canada
xiteng.liu@qualvisual.net

*Abstract*—We discuss the applicability of compressed sensing theory. We take a genuine look at both experimental results and theoretical works. We answer the following questions:
 1) What can compressed sensing really do?
 2) More importantly, why?

## I. WHAT CAN COMPRESSED SENSING DO?

Compressed sensing theory is described and studied as a panacea in many fields of science and engineering, evidenced by the website [2], which is built and managed by researchers of Rice University. We hereby take a genuine look at its performance from two most representative research results.

### A. Compressed Sensing Result of Rice University

The research result of Rice University on compressed sensing is called "single-pixel camera" and is posted at website [3]. It is marketed by the company InView Technology ([4]). Fig. 1 demonstrates the performance of compressed sensing result by Rice University in comparison with Rapid technology which implements system compression method. All original image data and the Rice results are from their website [3]. We can see that, the Rice results lose color and shape. In striking contrast, Rapid results miraculously achieve visually lossless to original images. In addition, Rapid runs hundreds times faster than Rice product. The Rapid demo software and measurement data (partial samples) can all be downloaded from website [1].

### B. Compressed Sensing Result of MIT

The research result of MIT (Massachusetts Institute of Technology) on compressed sensing has been reported by MIT News at least three times, see website [5]. Fig. 2 demonstrates the performance of compressed sensing result by MIT in comparison with Rapid technology which implements system compression method. All original images and MIT results are from MIT News website [5] and thesis report [6] (page 90, Appendix B). Rapid demo software and measurement data (partial samples) can all be downloaded at the website [1].

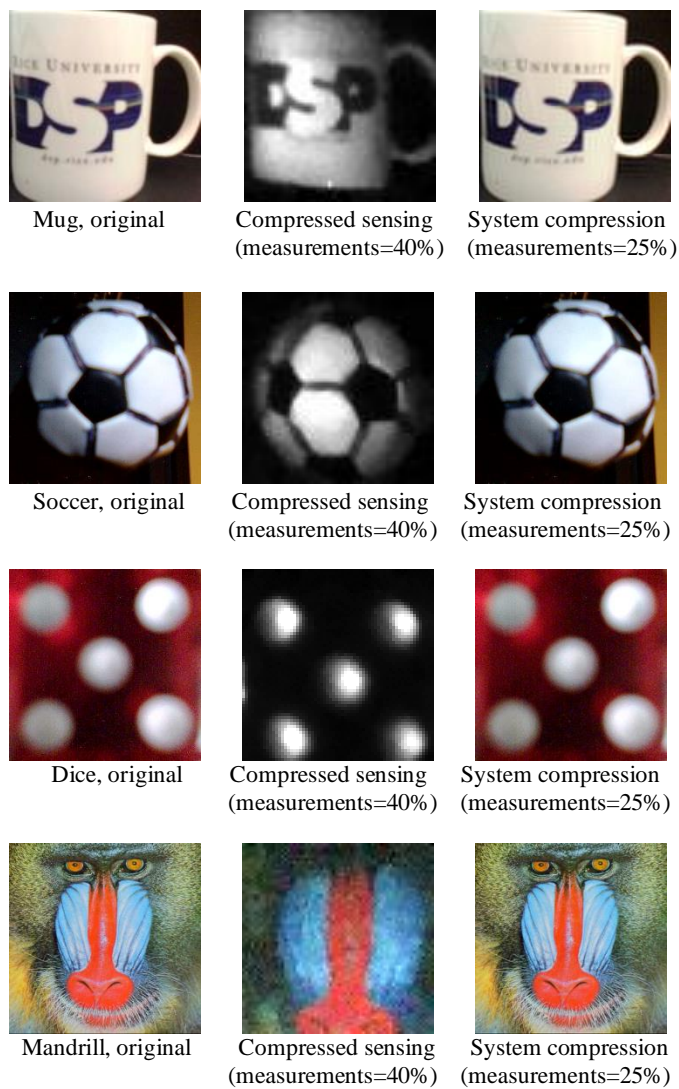

| Mug, original | Compressed sensing (measurements=40%) | System compression (measurements=25%) |
| Soccer, original | Compressed sensing (measurements=40%) | System compression (measurements=25%) |
| Dice, original | Compressed sensing (measurements=40%) | System compression (measurements=25%) |
| Mandrill, original | Compressed sensing (measurements=40%) | System compression (measurements=25%) |

**Fig. 1   Performance of compressed sensing by Rice University.**

## II. WHY COMPRESSED SENSING PERFORM POOR?

The poor performance of compressed sensing is caused by its lack of solid theoretical support. We hereby give detailed analysis over the mathematical framework of compressed sensing. We provide a counterexample which disproves the foundation theorem of compressed sensing theory. Furthermore, by a simple analysis, we can find that the main theoretical result which directs methodological practice of compressed sensing actually does not provide useful support for practical applications.

### A. Disprove the Fooundation Theorem

The foundation theorem of compressed sensing theory, Theorem 1.1 in [7] states that "Suppose that the signal length $N$ is a prime integer. Let $\Omega$ be a subset of $\{0, \ldots, N\text{-}1\}$, and let f be a vector supported on T such that $|T| \leq \frac{1}{2}|\Omega|$. Then f can be reconstructed uniquely from $\Omega$ and $\hat{f}$." It is further clarified that "Theorem 1.1 asserts that one can reconstruct f from 2|T| frequency samples (and that, in general, there is no hope to do so from fewer samples). In principle, we can recover f exactly by solving the combinatorial optimization problem

$$(P_0) \quad \min_{g \in C^N} ||g||_0, \quad \hat{g}|_\Omega = \hat{f}|_\Omega,$$

where $||g||_0$ is the number of nonzero terms $\#\{t, g(t) \neq 0\}$." This theorem is praised as "a very significant advance" in [10].

Since $N$ is a prime number, we know that $N = 2k+1$ for an integer k. Let $\Omega = \{0,1,3,\ldots,i,\ldots,N-i,\ldots,N-3,N-1\}$ such that it has symmetric structure for $1 \leq i \leq k$. Let $n = |\Omega|$. We can show that a vector f supported on T such that $|T| \leq \frac{1}{2}|\Omega|$ is not necessarily the unique solution of problem $(P_0)$ and hence cannot be reconstructed from $\Omega$ and $\hat{f}$.

Let $\Psi$ be the $n \times N$ matrix that contains the n rows of $N \times N$ Fourier matrix which are indexed by elements of $\Omega$,

$$\Psi = \frac{1}{\sqrt{N}} \begin{bmatrix} 1 & 1 & 1 & \cdots & 1 \\ 1 & \omega_N^1 & \omega_N^2 & \cdots & \omega_N^{N-1} \\ 1 & \omega_N^3 & \omega_N^6 & \cdots & \omega_N^{3(N-1)} \\ \vdots & \vdots & \vdots & \cdots & \vdots \\ 1 & \omega_N^{N-1} & \omega_N^{2(N-1)} & \cdots & \omega_N^{(N-1)(N-1)} \end{bmatrix},$$

where $\omega_N = e^{2\pi j/N}$, $j = \sqrt{-1}$. If any n columns of $\Psi$ are linearly independent, then $\Psi$ is said to be *maximally robust* and f is the unique solution of $(P_0)$. Otherwise, f can not be exactly recovered by solving $(P_0)$. In the

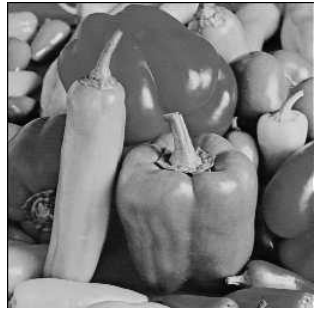
Peppers, original 256*256

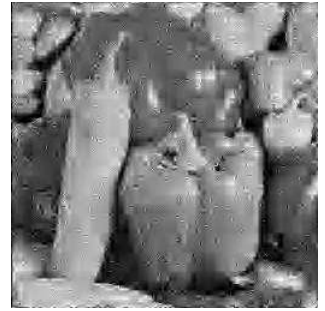
Compressed sensing (measurements = 17000)

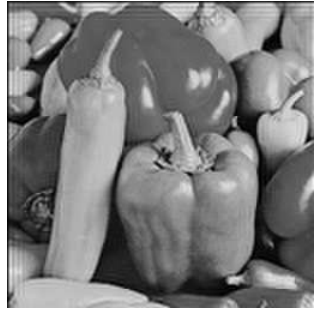
System compression (measurements = 17000)

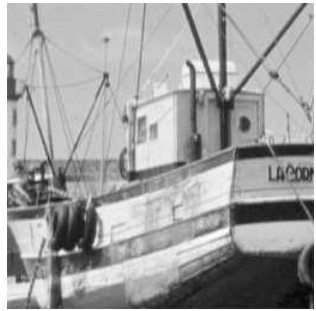
Boat, original 256*256

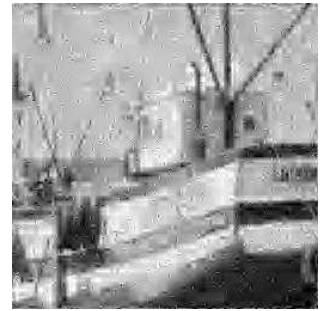
Compressed sensing (measurements = 25000)

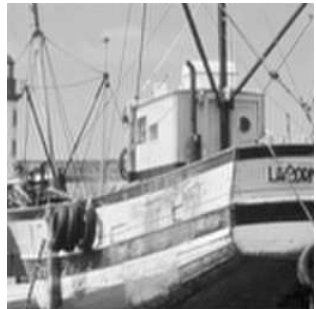
System compression (measurements = 17000)

**Fig. 2** Performance of compressed sensing by MIT.

followings, we shall show that $\Psi$ can not be maximally robust for any prime number $N$.

Define $N \times N$ matrix Q as
$$Q = \begin{bmatrix} 1 & 0 & 0 \\ 0 & \frac{1}{\sqrt{2}}I_k & \frac{1}{\sqrt{2}}J_k \\ 0 & \frac{j}{\sqrt{2}}I_k & -\frac{j}{\sqrt{2}}J_k \end{bmatrix}, \quad J_k = \begin{bmatrix} 0 & & 0 & 1 \\ 0 & & 1 & 0 \\ 0 & \ddots & & 0 \\ 1 & 0 & & \end{bmatrix},$$

and $I_k$ is the $k \times k$ identity matrix. Using this matrix we can get the matrix $\Phi$ by $\Phi = \Psi Q^*$,

$$\Phi = \sqrt{\frac{2}{N}} \begin{bmatrix} \sqrt{1/2} & 1 & \cdots & 1 & 0 & \cdots & 0 \\ \sqrt{1/2} & c(1) & \cdots & c(k) & s(1) & \cdots & s(k) \\ \sqrt{1/2} & c(3) & \cdots & c(3k) & s(3) & \cdots & s(3k) \\ \vdots & \vdots & \cdots & \vdots & \vdots & \cdots & \vdots \\ \sqrt{1/2} & c(t) & \cdots & c(kt) & s(t) & \cdots & s(kt) \end{bmatrix},$$

where $c(i) = \cos\left(\frac{2\pi i}{N}\right), s(i) = \sin(\frac{2\pi i}{N})$ and $t = N-1$. Matrix $C_2$ contains the constant column and all cosine columns of $\Phi$ and matrix $S_2$ contains all sine columns. Resulting from the symmetric structure of $\Omega$, $\Phi$ enjoys a symmetric structure: for $l = 1, \ldots, k$ and $i = 1, \ldots, k$ we have $c(i(N-l)) = c(il)$ and $s(i(N-l)) = -s(il)$. Using this symmetric structure, one can prove that the cosine and sine columns of $\Phi$ are orthogonal, i.e. $C_2 \perp S_2$. Since $\Psi$ is a partial matrix of a Fourier matrix, we have $rank(\Phi) = rank(\Psi) = n$ and
$$n = \begin{cases} k+1, & \text{if } k \text{ is even}; \\ k+2, & \text{otherwise}. \end{cases}$$
Hence we know for certain that $n \geq (k+1)$. Let $M = \Phi^T\Phi$, $M_1 = C_2^T C_2$ and $M_2 = S_2^T S_2$. Thanks to the symmetric structure and orthogonality between $C_2$ and $S_2$, we may get
$$M = \begin{bmatrix} M_1 & 0 \\ 0 & M_2 \end{bmatrix}.$$
Assume matrix $\Phi$ is maximally robust, i.e. any $n$ columns of $\Phi$ are linearly independent. Then, the $(k+1) \leq n$ columns of $C_2$ are linearly independent which means $rank(C_2) = k+1$. The singular values of matrix $M_1$ are squares of singular values of $C_2$. Thus, matrices $M_1$ and $C_2$ have the same number of nonzero singular values, which means $rank(M_1) = rank(C_2) = k+1$. Similar argument may show that $rank(S_2) = k$ and hence $rank(M_2) = k$. Therefore, we have
$rank(M) = rank(M_1) + rank(M_2) = 2k+1 = N$.
On the other hand, $rank(M) = rank(\Phi) = n < N$. Thus, we get contradiction and hence $\Phi$ can not be maximally robust for any $N = 2k+1$. Since Q is invertible and unitary, $\Psi$ can not be maximally robust for any prime number N ([9]). There f can not be uniquely reconstructed by solving problem ($P_0$), because it is not necessarily the uniquely sparsest solution.

*B. Unwrap the Main Theorem*

The main theorem of compressed sensing theory,

Theorem 1.3 in [7] was proved based on Theorem 1.1 which is disproved above. Its variant version Theorem 1 in [8] states that "Fix $f \in R^n$ and suppose that the coefficient sequence of $f$ in the basis $\Psi$ is S-sparse. Select $m$ measurements in the $\Phi$ domain uniformly at random. Then if

$$m \geq C \cdot \mu^2(\Phi, \Psi) \cdot S \cdot \log n \quad (1)$$

for some positive constant $C$, the solution to $l_1$-norm minimization problem is exact with overwhelming probability." Here $C = 46$ and $\mu(\Phi, \Psi) \in [1, \sqrt{n}]$.

Before applying this theorem to engineering practice, we **must** ask the question "What can it tell us?" We can answer this question with a very simple example. Let the length of signal $f$ be $n=1024$. We need to check how sparse $x$ can be to be suitable for application of this theorem, if we recover signal $f$ from its 50% measurements with "overwhelming probality" in an optimal situation when $\mu(\Phi, \Psi) = 1$. Now by (1) we get

$$S \leq \frac{m}{C \cdot \mu^2(\Phi,\Psi) \cdot \log n} = \frac{512}{46 \times \log 1024} \approx 1.6.$$

This means at most 2 of the 1024, or 0.2% coefficients in vector $x$ can be nonzero while all other 99.8% coefficients must be zero! In fact, this type of signals have no realistic applicability in engineering practices. In other words, this theorem factually delivers no valuable information for practical compressed sensing applications. Fig. 3 demonstrates what it means for a signal to have only 0.2% nonzero coefficients. By a simple case study, we sadly find that the main theorem of compressed sensing theory actually tells nothing valuable for practical designs.

III. CONCLUSION

A tree on rotten root cannot bear fruit. With a counterexample, we disproved the foundation theorem of compressed sensing theory. Furthermore, from a simple example, we find that the main theorem provides no genuine support for practical methods. Our works reveal the fundamental reason why existing compressed sensing methods do not perform as promised. Without solid theoretical support, compressed sensing methods unavoidably perform poor.

REFERENCES

[1] http://qualvisual.net
[2] http://dsp.rice.edu/cs
[3] http://dsp.rice.edu/cscamera
[4] http://inviewcorp.com
[5] http://web.mit.edu/newsoffice/topic/compressed-sensing.html
[6] Radu Berinde, "Advances in Sparse Signal Recovery Methods," MIT thesis, Aug 2009
[7] Cand'es E, Romberg J, Tao T., "Robust uncertainty principles: Exact signal reconstruction from highly incomplete frequency information", IEEE Transactions on Information Theory 2006; 52:489509.
[8] Emmanuel Candes and Michael Wakin, "An introduction to compressive sampling," IEEE Signal Processing Magzine, March 2008
[9] M. Pushel and J. Kovacevic, "Real, tight frames with maximal robustness to erasures," Proc. Data Compression Conference, March 2005
[10] D. L. Donoho. "Compressed sensing," IEEE Trans. Inform. Theory, 52(4):1289–1306, 2006.

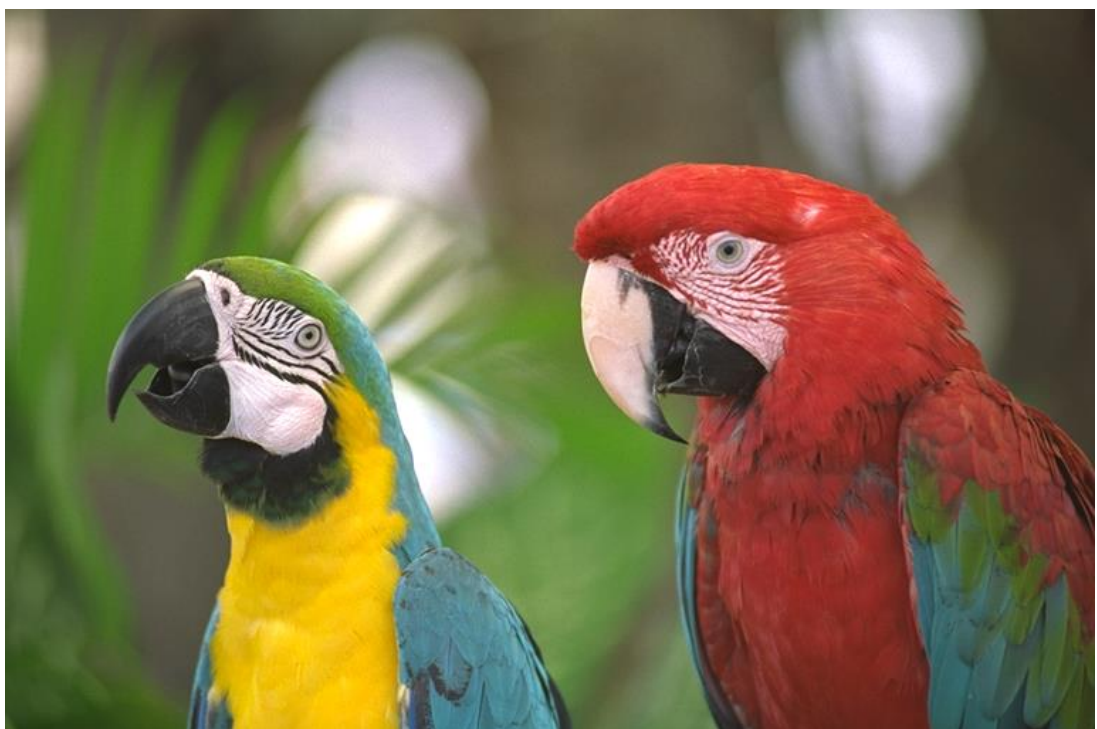

Original signal, 768*512*24

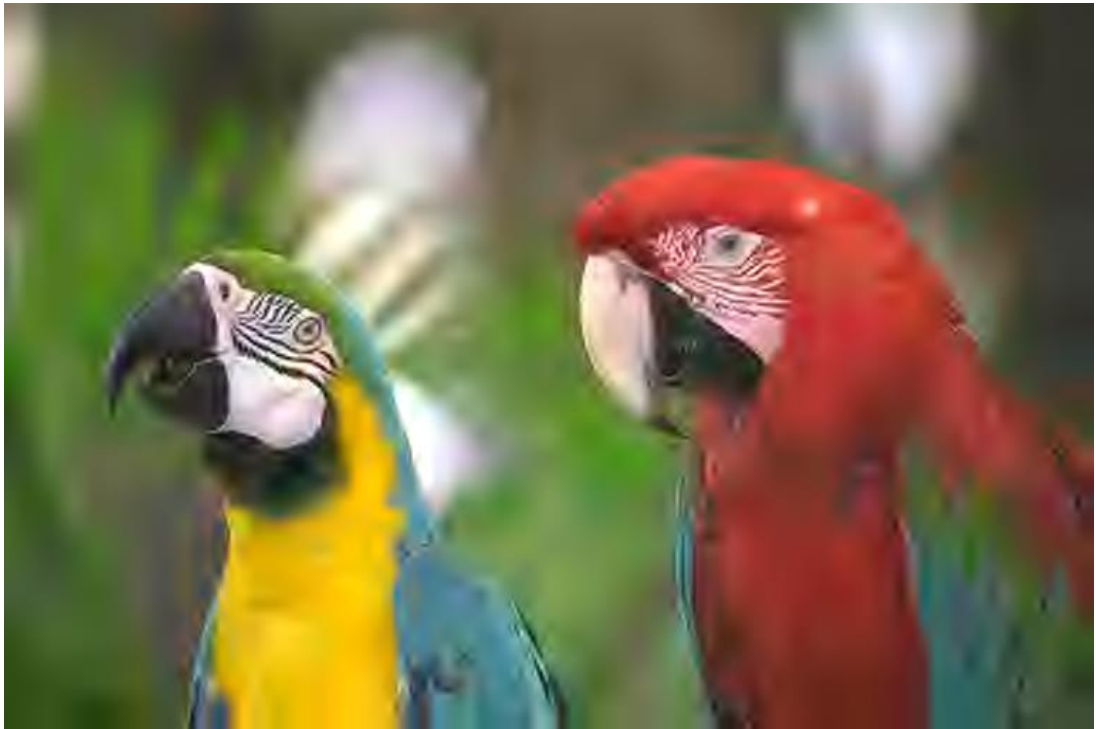
Signal with 0.2% nonzero wavelet coefficients

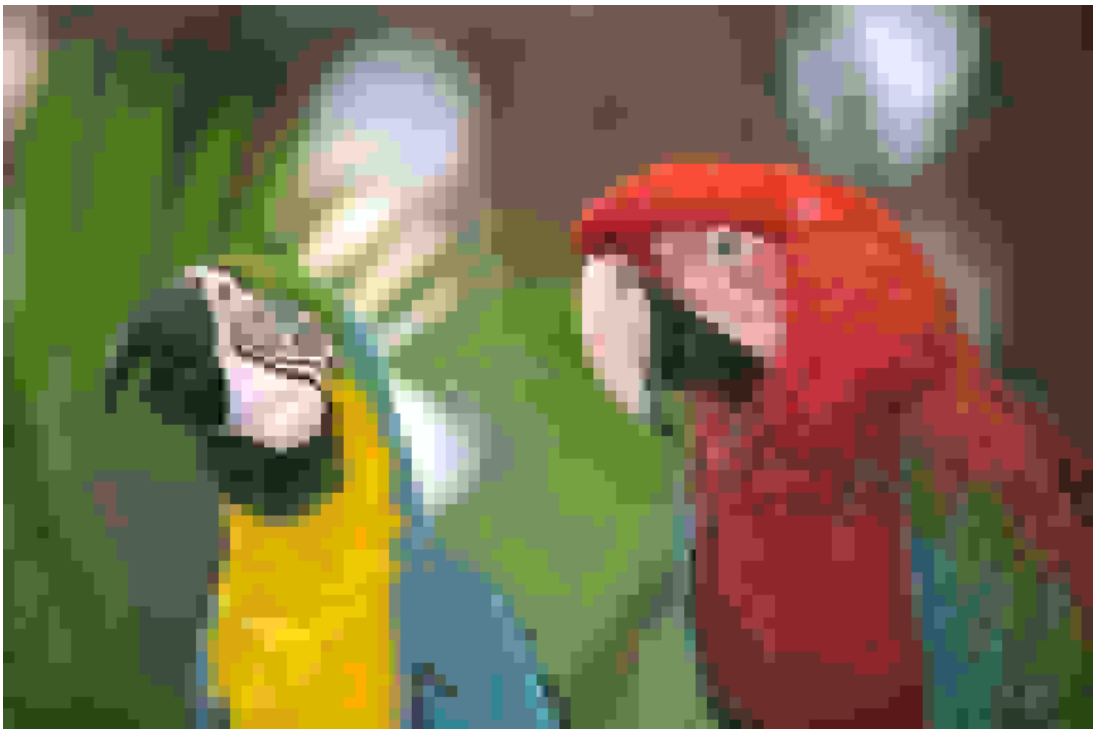
Signal with 0.2% nonzero DCT coefficients

**Fig. 3    Signal quality with only 0.2% nonzero coefficients.**